\documentclass[12 pt]{article} 
\usepackage{times,bm}
\usepackage[affil-it,]{authblk}
\usepackage[pagebackref=false,colorlinks,linkcolor=blue,citecolor=blue]{hyperref}
\usepackage{geometry}
\usepackage{graphicx}
\usepackage{cite}
\usepackage{amsmath}
\usepackage{amsfonts}
\usepackage{amssymb}
\usepackage{color}
\usepackage{float}
\usepackage{multirow,multicol}
\usepackage{tabulary}
\usepackage[flushleft]{threeparttable}
\usepackage{pgfplots,subfigure}
\usepackage{xcolor}
\numberwithin{equation}{section}
\usepackage{tikz}
\usepgflibrary{arrows}
\usetikzlibrary{shapes.callouts}
\tikzset{  
	level/.style   = { thick, },
	connect/.style = { dotted, red   },
	notice/.style  = { draw, rectangle callout, callout relative pointer={#1} },
	label/.style   = { text width=2cm }
}

\begin{document}

\title{ \emph{Observation of Ultra Fine Structures in Energy Levels of Prolate Nuclei}}

\author[1]{\textit{Hassan Hassanabadi}
\thanks{ Electronic address: hha1349@gmail.com  }}

\author[1]{\textit{Hadi Sobhani}
	\thanks{Electronic address: hadisobhani8637@gmail.com (Corresponding author)}}

\affil[1]{{\small Physics Department, Shahrood University of Technology, Shahrood, Iran \\ P. O. Box: 3619995161-316.}}

\maketitle

\begin{abstract}
  \textit{This article contains a discussion in which we showed that observation of splitting in the energy levels of prolate nuclei, is possible. Similar effects is atomic physics is known as Zeeman effect which is well-known, but in nuclear physics such topic has not been discussed and mentioned if it is possible or is not. In this article, after introducing deformation in commutation relation in three dimensional, these relations has been used in X(3) model. After enough derivation, we evaluate energy spectrum relation for considered system which has splitting. Because of considerations during the derivation, we'd like to call such effect as ultra fine structures in energy levels. At the end some plots have been depicted which illustrate the results. }
\end{abstract}

\begin{small}
\end{small}

\begin{small}
\end{small}

\newpage
\section{Introduction}

Investigation of Bohr Hamiltonian \cite{1} has been revived in special solutions \cite{2,3} because the advent of critical point symmetries \cite{4,5,6,7}, manifested experimentally \cite{8,9,10} in nuclei on or near the point of shape phase transitions. However, phase transitions in nuclei have been originally found out \cite{7,11} in Interacting Boson Model \cite{12}. As first examples of critical point symmetries E(5) symmetry \cite{7} and the X(5) symmetry \cite{8} can be mentioned that the first one corresponds to the second order phase transition between spherical and $\gamma$-unstable (soft with respect to axial asymmetry) nuclei and the second one  is appropriate for the first order phase transition between spherical and prolate deformed nuclei. After the critical point symmetries were proposed, another movement began to start and develop but this time, the goal was supposing rigidity for $\gamma$ collective parameter to new features for the field of the nuclear shape phase transitions. Advances of this thinking  resulted in Z(4) \cite{13} and X(3) \cite{14} models as $\gamma$-rigid counterparts of the Z(5) and X(5) models. Z(4) is described by the Davydov-Chaban Hamiltonian \cite{15} which emerges from the Bohr Hamiltonian by freezing the $\gamma$ degree of freedom. Going further and imposing axial symmetry, one obtains after quantization in the remaining curvilinear coordinates a new Hamiltonian \cite{14} associated to the X(3). At the end of this part of introduction we refer reader to Refs. \cite{1,2,3,4,5,6,7,8,9,10,11,12,13,14,15} 
\cite{16,17,18,19,20} 
to study more about above models.

In the last decades, non-commutative theories have attracted 
many interests and efforts in different areas such as  geometry \cite{21}, condensed matter \cite{22,23,24} and quantum gravity
both from the theoretical \cite{25,26} and phenomenological \cite{27,28,29} point of view. The most important motivations for studying NC spaces comes from string theory. The non-commutative concepts in physics can be traced back to the momentum algebra. These concepts are upgraded to any pair of conjugate variables, called Heisenberg’s algebra in quantum mechanics. It implies that the phase space of elementary particles has an intrinsic uncertainty and particles have nonlocal behavior in the microscopic world \cite{30,31,32}.

In this article we are using non-commutative theories in X(3) models to find out what will happen if effects of non-commutativity is considered in X(3) model. It will shown that interesting results will obtain which has not been appeared. We organized this article as follows: Sec. \ref{Sec. 2} includes basic theories of non-commutative pas space in three dimensional. In Sec. \ref{Sec. 3} after describing X(3) model, effects of non-commutativity has been considered then the results will evaluated. at the end numerical results are shown to illustrate the results.

\section{Non-Commutative Phase Space in Three Dimensional}\label{Sec. 2}
In ordinary three dimensional commutative space, we face with below relations \cite{33}

\begin{align}\label{2-1}
\left[ {{x_i},{x_j}} \right] = 0, \quad \left[ {{p_i},{p_j}} \right] = 0, \quad \left[ {{x_i},{p_j}} \right] = i\hbar {\delta _{ij}}, \quad \left( {i,j = 1,2,3} \right).
\end{align}
It is assumed that at very tiny scale, such as string scales, space coordinates may not have such commutative relations. Throughout of this article we distinguish non-commutative operators with hat symbol. Commutative relations between coordinate and momentum in three dimensional non-commutative phase are \cite{33}
\begin{align}\label{2-2}
\left[ {{{\hat x}_i},{{\hat x}_j}} \right] = i{\epsilon _{ijk}}{\vartheta _k},\quad \left[ {{{\hat p}_i},{{\hat p}_j}} \right] = i{\epsilon _{ijk}}{{\bar \vartheta }_k}, \quad \left[ {{{\hat x}_i},{{\hat p}_j}} \right] = i\hbar {\delta _{ij}},\quad \left( {i,j = 1,2,3} \right)
\end{align}    
where $\epsilon_{ijk}$ is used as Levi-Civita symbol and the summation convention is supposed as well as $\vartheta$ and $\bar{\vartheta}$ are constant. These spaces can be connected to each other by considering 
\begin{align}
\label{2-3}
{{\hat x}_i} = {x_i} + \frac{{{\epsilon _{ijk}}{\vartheta _j}}}{{2{\hbar _{eff}}}}{p_k},  \\
\label{2-4}
{{\hat p}_i} = {p_i} - \frac{{{\epsilon _{ijk}}{{\bar \vartheta }_j}}}{{2{\hbar _{eff}}}}{x_k},  \\
\label{2-5}
{\hbar _{eff}} = \hbar \left( {1 + \frac{{\vartheta \bar \vartheta }}{{2{\hbar ^2}}}} \right).  
\end{align} 
According Eqs.\eqref{2-3} and \eqref{2-4}, squared form of coordinate and momentum in non-commutative can be derived as
\begin{align}
\label{2-6}
\hat x_i^2 = {r^2} - \frac{{\vartheta .L}}{\hbar } + \frac{1}{{4{\hbar ^2}}}\left[ {{\vartheta ^2}{p^2} - {{\left( {\vartheta .p} \right)}^2}} \right],  \\
\label{2-7}
\hat p_i^2 = {p^2} - \frac{{\bar \vartheta .L}}{\hbar } + \frac{1}{{4{\hbar ^2}}}\left[ {{{\bar \vartheta }^2}{r^2} - {{\left( {\bar \vartheta .r} \right)}^2}} \right], 
\end{align} 
where
\begin{equation}
\begin{gathered}
\vartheta .L = {\vartheta _i}{L_i}, \quad \bar \vartheta .L = {{\bar \vartheta }_i}{L_i} \quad ,{\vartheta ^2} = {\vartheta _i}{\vartheta _i}, \hfill \\
{{\bar \vartheta }^2} = {{\bar \vartheta }_i}{{\bar \vartheta }_i}, \quad {r^2} = {x_i}{x_i}, \quad {p^2} = {p_i}{p_i}, \hfill \\
{L_i} = {\epsilon _{ijk}}{x_j}{p_{k.}}. \hfill \\ 
\end{gathered} 
\end{equation}

In the next section, this space will be used instead of ordinary space.
\section{Description of X(3) Model}
\label{Sec. 3}
In the collective model which Bohr \cite{1} discussed, the classical expression of the kinetic energy corresponding to $\beta$ and $\gamma$ vibrations ($\beta$ and $\gamma$ are usual collective parameter ) of the nuclear surface and rotation of the nucleus can be written as  \cite{14,17}
\begin{equation}\label{3-1}
T = \frac{1}{2}\sum\limits_{k = 1}^3 {{J_k}{w'_k}^{2} + \frac{B}{2}\left({{\dot \beta }^2} + {{\left( {\beta \dot \gamma } \right)}^2}\right)} 
\end{equation}   
in which $B$ is the mass parameter and $J_k$, the three principal irrotational moments of inertia, has the form 
\begin{equation}\label{3-2}
{J_k} = 4B{\beta ^2}{\sin ^2}\left(\gamma  - \frac{{2\pi }}{3}k\right)
\end{equation}
and the components of the angular velocity on the body-fixed k-axes, is shown by $w'_k$,$(k=1,2,3)$, which can be written in terms of the time derivatives of the Euler angles $\dot \phi ,\dot \theta ,\dot \psi $ 
\begin{equation}\label{3-3}
\begin{gathered}
{{w'}_1} = \dot \phi \sin \theta \cos \psi  + \dot \theta \sin \psi , \hfill \\
{{w'}_2} = \dot \phi \sin \theta \sin \psi  + \dot \theta \cos \psi , \hfill \\
{{w'}_3} = \dot \phi \cos \theta  + \dot \psi . \hfill \\ 
\end{gathered} 
\end{equation}
In the manner that Davydov-Chaban adopted the nucleus is rigid, it means $\dot \gamma  = 0$, as well as there is a special consideration that being the axially symmetric prolate case of $\gamma =0$. Equaling
$J_1=J_2=3B{\beta}^2$, results vanishing the third irrotational moment of inertia $J_3$. In this case the kinetic energy of Eq. (\ref{3-1}) becomes 
\begin{align}\label{3-4}
T &= \frac{3}{2}B{\beta ^2}\left({w'_1}^2 + {w'_2}^2\right) + \frac{B}{2}{\beta ^2} \nonumber\\
&= \frac{B}{2}\left[3{\beta ^2}\left({{\dot \phi }^2}{\sin ^2}\theta  + {{\dot \theta }^2}\right) + {{\dot \beta }^2}\right].  
\end{align} 
As it can be understood easily that there is only three degree of freedom in this case. Considering the generalized coordinates ${q_1}=\phi$, ${q_2}=\theta$, and ${q_3}=\beta$, the kinetic energy becomes a quadratic form of the time derivatives of the generalized coordinates 
\begin{equation}\label{3-5}
T = \frac{B}{2}\sum\limits_{i,j = 1}^3 {{g_{ij}}{{\dot q}_i}{{\dot q}_j}} ,
\end{equation}
although in Bohr Hamiltonian the matrix $g_{ij}$ has non-diagonal and 5-dimensional form, but here it is a diagonal matrix as
\begin{equation}\label{3-6}
{g_{ij}} = \left( {\begin{array}{*{20}{c}}
	{3{\beta ^2}{{\sin }^2}\theta }&0&0 \\ 
	0&{3{\beta ^2}}&0 \\ 
	0&0&1 
	\end{array}} \right).
\end{equation} 
Following the general procedure of quantization in curvilinear coordinates one obtains the Hamiltonian operator
\begin{equation}\label{3-7}
\begin{gathered}
H = \frac{{{p^2}}}{{2B}} + U\left( \beta  \right) \hfill \\
H = \frac{{ - {\hbar ^2}}}{{2B}}\left[ {\frac{1}{{{\beta ^2}}}\frac{\partial }{{\partial \beta }}{\beta ^2}\frac{\partial }{{\partial \beta }} + \frac{{{\Delta _\Omega }}}{{3{\beta ^2}}}} \right] + U\left( \beta  \right), \hfill \\ 
\end{gathered} 
\end{equation}
in which $B$ is the mass parameter, $\beta$ is the collective variable and the angular part of Laplace operator is shown by
\begin{equation}
\label{3-8}
{\Delta _\Omega } = \frac{1}{{\sin \theta }}\frac{\partial }{{\partial \theta }}\sin \theta \frac{\partial }{{\partial \theta }} + \frac{1}{{{{\sin }^2}\theta }}\frac{{{\partial ^2}}}{{\partial {\phi ^2}}}.
\end{equation}

If we consider non-commutative phase space for our system, Eq.\eqref{3-7} should be rewrite as
\begin{align}
\label{3-9}
H = \frac{{{{\hat p}^2}}}{{2B}} + U\left( \beta  \right).
\end{align}
Considering potential well interaction of system
\begin{equation}
\label{3-10}
U\left( \beta \right) = \left\{ \begin{gathered}
0, \quad \beta  \leqslant {\beta _w}  \\
\infty , \quad \beta  > {\beta _w} \hfill \\ 
\end{gathered}  \right.
\end{equation}
where $\beta _w$ is width well, we obtain the Hamiltonian using Eq. \eqref{2-7} as
\begin{align}
\label{3-11}
&H = {H_0} + {H_{per}},  \\
\label{3-12}
&{H_0} = \frac{{{p^2}}}{{2B}},  \\
\label{3-13}
&{H_{per}} = \frac{{ - \bar \vartheta .L}}{{2B\hbar}},  
\end{align}
where we neglected terms $O\left( \bar{\vartheta}^2\right) $. It should be note that since $\bar{\vartheta}$ can admit small values, we consider the term including $\bar{\vartheta}$ parameter as perturbation of system.
We first derive wave function corresponding Eq. \eqref{3-12}. Assuming wave function as $\Psi \left( {\beta ,\theta ,\varphi } \right) = \frac{{f\left( \beta  \right)}}{{\sqrt \beta  }}{Y_{Lm}}\left( {\theta ,\varphi } \right)$, we get to spherical Bessel differential equation
\begin{align}
\label{3-14}
&\left[ {\frac{{{d^2}}}{{d{\beta ^2}}} + \frac{2}{\beta }\frac{d}{{d\beta }} + \left( {{\kappa ^2} - \frac{{{\nu ^2}}}{{{\beta ^2}}}} \right)} \right]f\left( \beta  \right) = 0,  \\
\label{3-15}
&{\kappa ^2} = \varepsilon  = \frac{{2BE}}{{{\hbar ^2}}} \\
\label{3-16}
&\nu  = \sqrt {\frac{{L\left( {L + 1} \right)}}{3} + \frac{1}{4}}, 
\end{align}
where the boundary condition is $f\left( {{\beta _w}} \right) = 0$. Solution of Eq. \eqref{3-14} can be written as 
\begin{align}
\label{3-17}
&f\left( \beta  \right) = \sqrt {\frac{2}{{\beta _w^2J_{\nu  + 1}^2\left( {{x_{s,\nu }}} \right)}}} {J_\nu }\left( {{\kappa _{s,\nu }}\beta } \right),  \\
\label{3-18}
&{\kappa _{s,\nu }} = \frac{{{x_{s,\nu }}}}{{{\beta _w}}},  
\end{align}
in which $x_{s,\nu }$ is is the s-th zero of the spherical Bessel function of the first kind ${J_\nu }\left( {{\kappa _{s,\nu }}{\beta _w}} \right)$.The corresponding spectrum of energy is in form of
\begin{align}
\label{3-19}
{E_{s,l}} = \frac{{{\hbar ^2}}}{{2B}}\kappa _{s,\nu }^2 = \frac{{{\hbar ^2}}}{{2B}}{\left( {\frac{{{x_{s,\nu }}}}{{{\beta _w}}}} \right)^2}.
\end{align}

For Hamiltonian \eqref{3-13}, we should evaluate expectation value $\left\langle {{H_{per}}} \right\rangle $ as first order of shifted energy. With the aim of Ladder operator ${L_ \pm } = {L_1} \pm i{L_2}$, we have
\begin{align}
\label{3-20}
\bar \vartheta .L = \left( {\frac{{{{\bar \vartheta }_1}}}{2} - i\frac{{{{\bar \vartheta }_2}}}{2}} \right){L_ + } + \left( {\frac{{{{\bar \vartheta }_1}}}{2} + i\frac{{{{\bar \vartheta }_2}}}{2}} \right){L_ - } + {{\bar \vartheta }_3}{L_3}.
\end{align}
By evaluating expectation value for $\bar \vartheta .L$, we obtain
\begin{align}
\label{3-21}
&\left\langle {{L_ \pm }} \right\rangle  = 0,  \\
\label{3-22}
&\left\langle {{{\bar \vartheta }_3}{L_3}} \right\rangle  = {{\bar \vartheta }_3}m\hbar , \quad - L < m < L. 
\end{align}
Therefore, Final form for energy spectrum is
\begin{align}
\label{3-23}
E = \frac{{{\hbar ^2}}}{{2B}}{\left( {\frac{{{x_{s,\nu }}}}{{{\beta _w}}}} \right)^2} - \frac{{m\bar \vartheta }}{{2B}},
\end{align}
where existence of $m$ in energy spectrum produces splitting in each energy level. This point is discussed numerically in following. Without lost of generality, we set $B=\hbar=\beta_w=1$. Setting $\bar{\vartheta}=0.1$, we have depicted energy spectrum for different angular momentum. Since $m$ is appeared in the spectrum, we face with splitting in energy levels. Since effects of $\bar{\vartheta}$ is so small, we call these splittings as ultra fine structures. In Fig. 1 these splitting is shown well. Also treatments of each splitting in terms of $\bar{\vartheta}$ changing is depicted in Fig. 2.  Since splitting has direct proportionality with $\bar{\vartheta}$, as $\bar{\vartheta}$ grows up, energy level splitting can be seen better.

\newpage
\section{Conclusion}

In this article, we considered non-commutative phase space in three dimensional. In such formalism, ordinary commutation relation between coordinate and momentum were changed and some of quantities, were derived. After presenting the X(3) model, we rewrote this model in the three dimensional non-commutative phase space. Effects of considered non-commutative phase space, were found in the energy eigen value relation which has dependence  on the projection of angular momentum.  Considering the angular momentum of the excited level $L$ of energy, we face with $2L+1$ splitting in the energy level of that excited state. We called such splitting the energy levels as ultra fine structures.  Obviously, observation of such effect is not as easy as the similar effect in the atomic physics, the Zeeman effect. Because the  ultra fine structures is due to non-commutative phase space, it can not be happened at the atomic scale. But our calculation showed existence of phenomena.




\begin{figure}[H]
	\label{Fig. 1}
	\centering
	\includegraphics[scale=0.5]{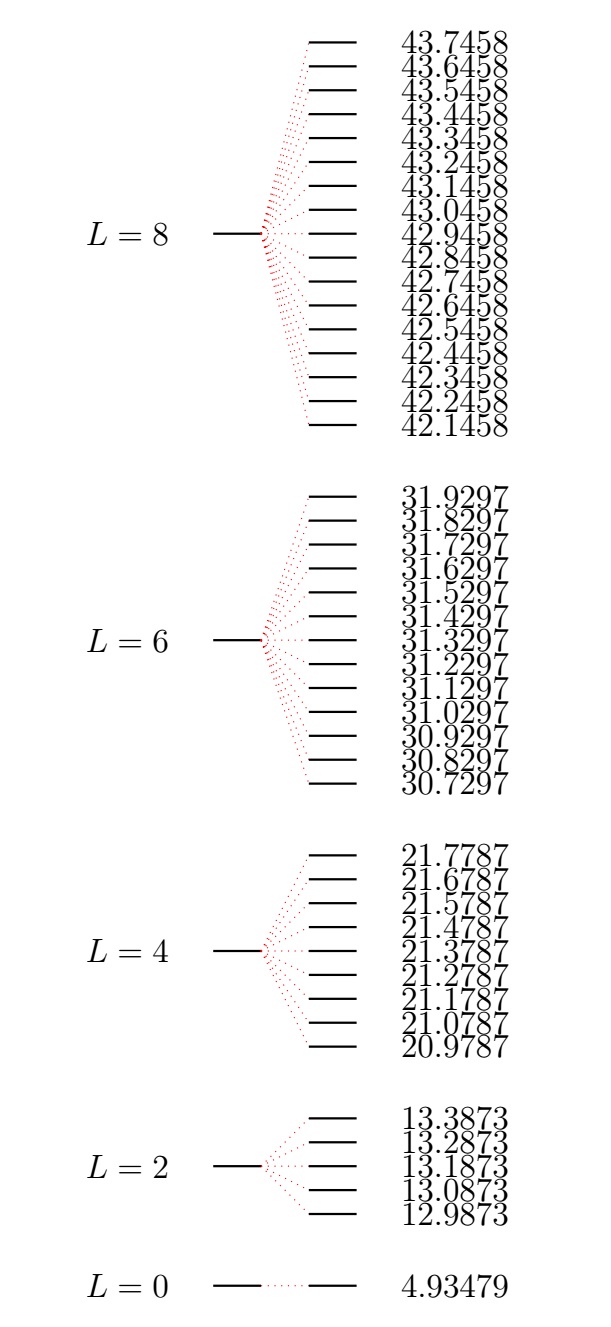}
	\caption{Observation of energy splitting for each value of angular momentum.}
\end{figure}

\begin{figure}[H]
	\label{Fig. 2}
	\centering
	\includegraphics[scale=0.4]{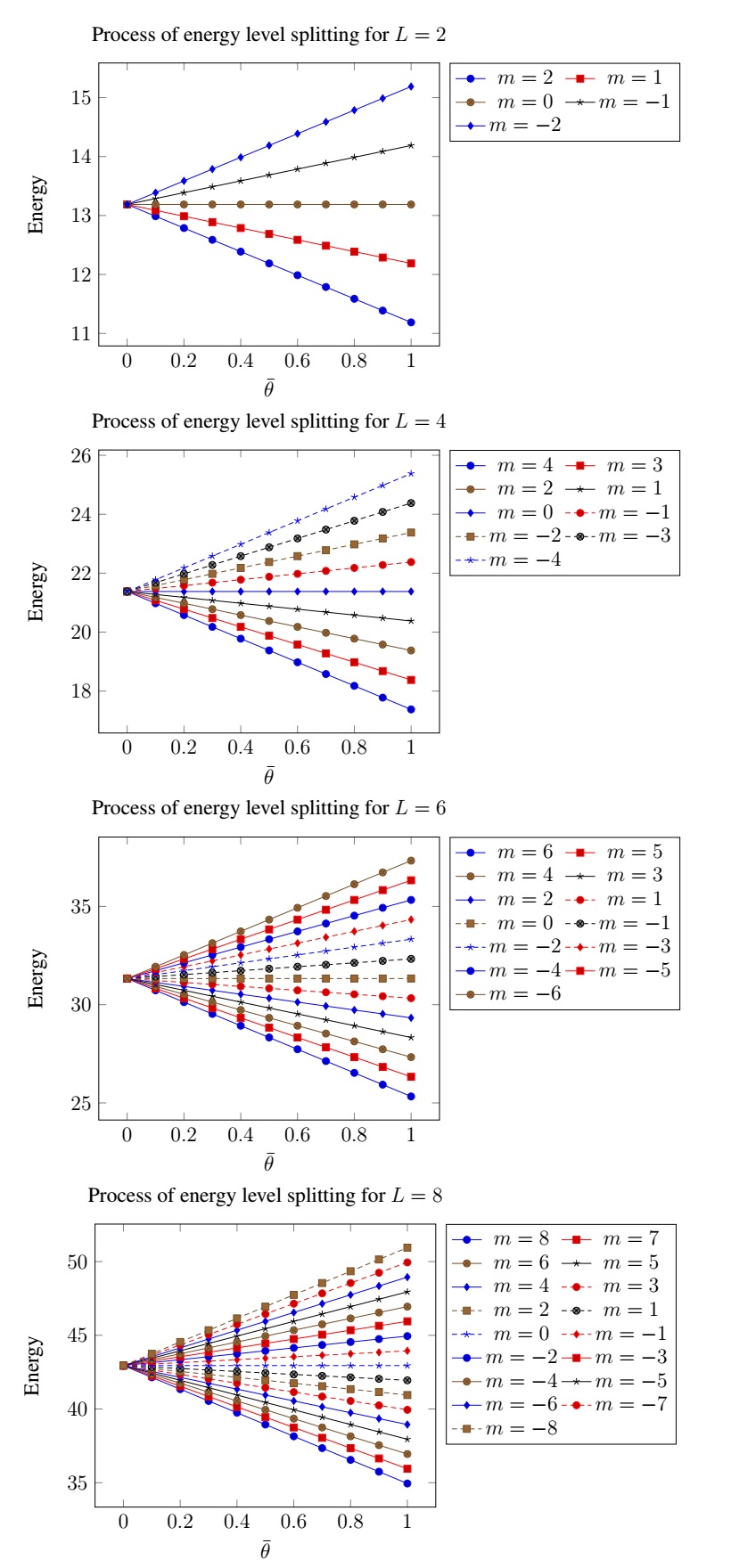}
	\caption{Effects of $\bar{\theta}$ on splitting of energy levels.}
\end{figure}

\begin{thebibliography}{99}

\bibitem{1}
A. Bohr, Mat. Fys. Medd. K. Dan. Vidensk. Selsk. 26,
no. 14 (1952).

\bibitem{2}
L. Fortunato, Eur. Phys. J. A 26, s01, 1 (2005).

\bibitem{3}
R. F. Casten, Nat. Phys. 2, 811 (2006).

\bibitem{4}
F. Iachello, Phys. Rev. Lett. 85, 3580 (2000).

\bibitem{5}
F. Iachello, Phys. Rev. Lett. 87, 052502 (2001).

\bibitem{6}
R. F. Casten, Prog. Part. Nucl. Phys. 62, 183 (2009).
\bibitem{7}
P. Cejnar and J. Jolie, Prog. Part. Nucl. Phys. 62, 210
(2009).

\bibitem{8}
R. F. Casten and N. V. Zamfir, Phys. Rev. Lett. 85, 3584
(2000).

\bibitem{9}
R. F. Casten and N. V. Zamfir, Phys. Rev. Lett. 87,
052503 (2001).

\bibitem{10}
R. F. Casten and E. A. McCutchan, J. Phys. G: Nucl.
Part. Phys. 34, R285 (2007).

\bibitem{11}
D. H. Feng, R. Gilmore, and S. R. Deans, Phys. Rev. C
23, 1254 (1981).

\bibitem{12}
F. Iachello and A. Arima, The Interacting Boson Model
(Cambridge University Press, Cambridge, 1987).

\bibitem{13}
D. Bonatsos, D. Lenis, D. Petrellis, P. A. Terziev and I. Yigitoglu,  Phys. Lett. B 621 102 (2005).

\bibitem{14}
D. Bonatsos, D. Lenis, D. Petrellis, P. A. Terziev and I. Yigitoglu   Phys. Lett. B 632 238 (2006).

\bibitem{15} 
A. S. Davydov  and A. A. Chaban,  Nuclear Physics 20 499 (1960).

\bibitem{16}
L. Naderi, H. Hassanabadi and H. Sobhani, Int, J, Mod. Phys. E, 25, 1650029  (2016).

\bibitem{17}
H. Sobhani and H. Hassanabadi, Phys. Lett. B, 760, 1 (2016).

\bibitem{18}
H. Sobhani and H. Hassanabadi, Nucl. Phys. A, 957, 177 (2016).

\bibitem{19}
H. Sobhani and  H. Hassanabadi, Int, J, Mod. Phys. E, 25, 1650073 (2016).

\bibitem{20}
H. Sobhani and  H. Hassanabadi, Mod. Phys. Lett. A, 31, 1650152 (2016).

\bibitem{21}
A. Connes, “Noncommutative geometry”, Academic Press, London and San Diego (1994).

\bibitem{22}
O. F. Dayi and A. Jellal, Phys. Lett. A 287 , 349 (2001). 

\bibitem{23}
J. Bellissard, A. van Elst and H. Schulz-Baldes, J. Math. Phys. 35 , 5373 (1994).

\bibitem{24}
H. Falomir, J. Gamboa, M. Loewe and M. Nieto, J. Phys. A, 45,  135308 (2012).

\bibitem{25}
N. Seiberg and E. Witten,  JHEP 9909 , 032 (1999).

\bibitem{26}
E. Witten,  Nucl. Phys. B 268 (1986).

\bibitem{27}
 I. Hinchliffe, N. Kersting and Y. L. Ma, Int. J. Mod. Phys. A 19 , 179 (2004).

\bibitem{28}
 G. Amelino-Camelia,  Living Rev. Rel. 16 , 5 (2013)

\bibitem{29}
 H. Falomir, S. A. Franchino Vias, P. A. G. Pisani and F. Vega,  JHEP 1312 , 024 (2013).

\bibitem{30}
M. R. Douglas and N. A. Nekrasov, Rev. Mod. Phys. 73, 977 (2001).

\bibitem{31} 
R. J. Szabo, Phys. Rep. 378, 207 (2003).

\bibitem{32}
H. Snyder, Phys. Rev. 71, 38 (1947).

\bibitem{33}
W. S. Chung, DOI: 10.13140/RG.2.1.1237.7368 (2016).

\end{thebibliography}
\end{document}